\documentclass[twocolumn,english]{revtex4-1}
\usepackage[T1]{fontenc}
\usepackage{amstext,graphicx,amsmath}
\usepackage{babel}
\begin{document}

\title{Macroscopic current generated by local division and apoptosis in a
minimal model of tissue dynamics}

\author{Emma Mitchell$^{\text{a}}$ and Elsen Tjhung$^{\text{a}}$}

\affiliation{$^{\text{a}}$Department of Physics, University of Durham, Science
Laboratories, South Road, Durham DH1 3LE, UK; E-mail: elsen.tjhung@durham.ac.uk}

\date{09/28/21}

\begin{abstract}
We consider a minimal computational model of tissue dynamics with
two active ingredients: local particle division and apoptosis. We
neglect other non-equilibrium effects such as self-propulsion. We
simulated the steady state dynamics inside an asymmetric channel and
we found a net macroscopic current along the channel. Although such
macroscopic current in a similar geometry has been detected in swimming
bacteria, our results showed that local division and apoptosis are
sufficient to generate a macroscopic current, without any need for
a self-propulsion/swimming mechanism. Our results might have applications
in tissue engineering such as controlling tissue growth via a geometrically
non-uniform substrate.

\end{abstract}
\maketitle

\section{Introduction}

\begin{figure}
\centering
  \includegraphics[width=0.8\columnwidth]{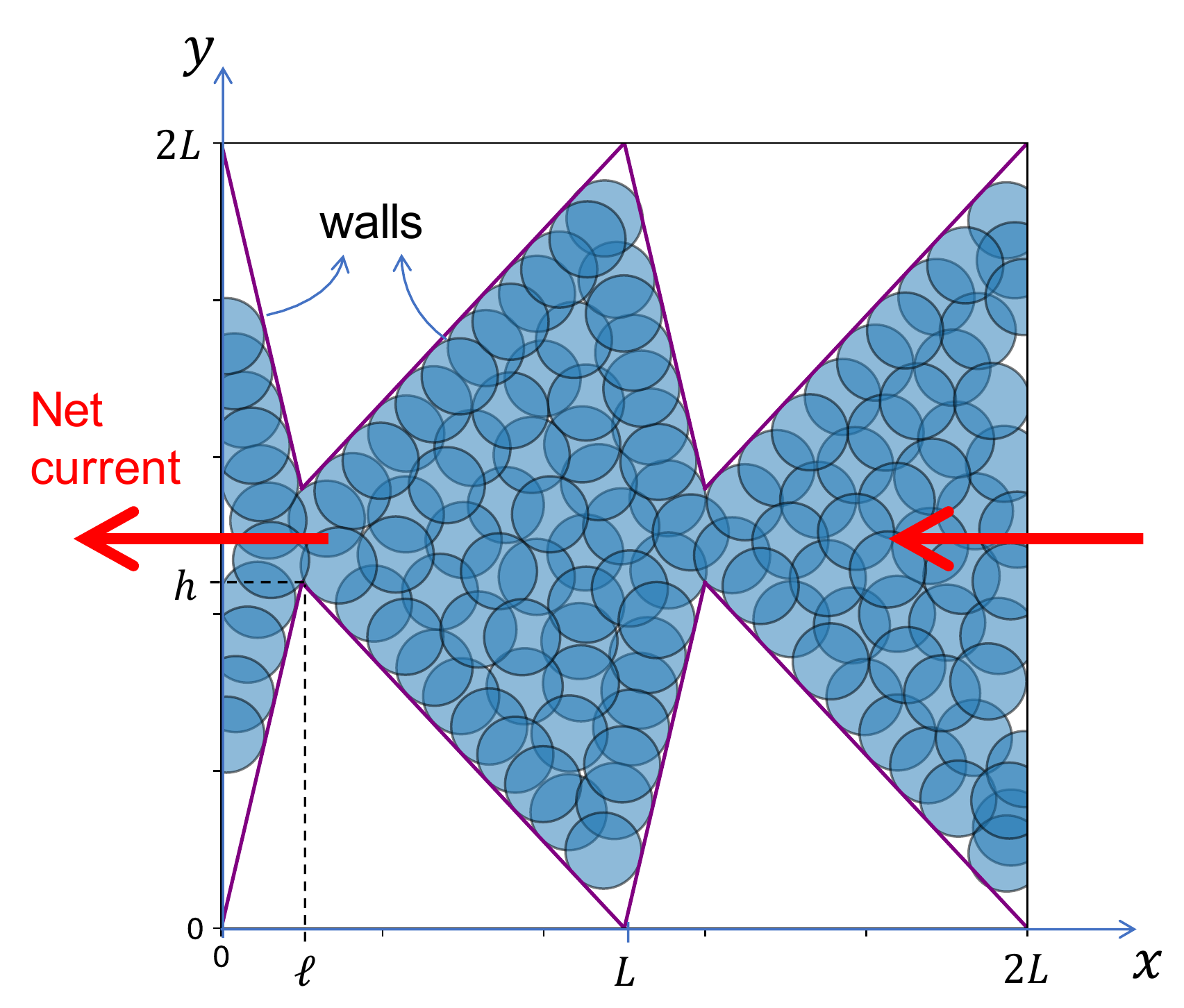}
  \caption{
  The geometry considered in our model.
  We consider a two-dimensional box of size $2L\times2L$ with periodic boundary conditions in the $x$-direction.
  Along $y$-direction, the particles are confined in between two corrugated walls with periodicity $L$, peak height $h$ and skewness $\ell\in[0,L]$.}
  \label{fig:geometry}
\end{figure}

It has been shown that swimming bacteria have the ability to perform useful macroscopic work.
For example, when we place an asymmetric cog inside a bath full of swimming bacteria,
the bacteria can rotate the cog in one direction persistently~\cite{DiLeonardo-2010-ratchet} (the size of the cog is around $30$ times larger than the bacteria).
Individually, the bacteria swim in a completely random direction.
However, interactions between the bacteria and the surface of the cog can break time reversal symmetry, giving rise to a macroscopic current circling around the cog.
This feature is of course not possible in an equilibrium system like passive Brownian particles.
Another manifestation of the same phenomenon happens when we place swimming bacteria inside a long asymmetric pipe (similar to one depicted in Fig.~\ref{fig:geometry}).
Experiments~\cite{Galajda-2007-ratchet} showed that the asymmetry of the pipe induced the bacteria to move, on average, in one direction.
Furthermore, it has also been shown theoretically~\cite{Lecomte-2020-ratchet} and numerically~\cite{Stenhammar-2016-ratchet,Reichhardt-2017-ratchet,Reichhardt-2013-ratchet} that the random swimming direction (for any generic micro-swimmers) can be rectified through the presence of an asymmetric potential.
This result is independent of any hydrodynamic or alignment interaction.

Another route to a large macroscopic current in swimming bacteria is through hydrodynamics.
The rotation of the flagella around the bacterium stirs the fluid around it and a spontaneous symmetry breaking
in the average orientations of the bacteria can give rise to a macroscopic flow (spontaneous flow transition~\cite{Voituriez-2005}).
This can happen in a symmetric/asymmetric potential, 
although the mechanism relies on hydrodynamic and alignment interactions~\cite{Markovich-2019-polar,Loisy-2018-rheology-polar}.
Such fluids are often called active extensile or contractile fluids~\cite{Marchetti-2013-review}.

%The above examples suggest that a brain is not always needed for a living organism to perform useful macroscopic work, as long as the organism is motile
%(or stirs the surrounding fluid and aligns with the neighbours).
In this paper, we consider a minimal computational model of living tissues which can divide (mitosis) or die (apotosis).
We show that local cellular division and apoptosis are sufficient to give rise to a macroscopic current inside an asymmetric channel,
without any recourse to self-propulsion/swimming (see Fig.~\ref{fig:geometry} and \ref{fig:vx}).
Since division and apoptosis are universal properties of all living systems, 
our results indicate that the ability for all living systems to perform useful macroscopic work is probably universal as well.
Finally, our results might also have applications in tissue engineering, for example, using ratchet-shaped scaffold or substrate to speed up tissue growth in one direction.

\section{Computational Model}

We consider a minimal computational model of living tissues in two dimensions with two key active ingredients: division and apoptosis.
We shall neglect hydrodynamic interactions and any other non-equilibrium process such as self-propulsion.
The aim of this paper is to demonstrate whether we can get a macroscopic current through rectification of local division and apoptosis alone.

% geometry
We approximate the cells as soft circular particles which can divide into two daughter particles or die (\emph{i.e.} removed from the system)~\cite{Barrat-2017-tissue,Henkes-2020-tissue,Tjhung-Berthier-2020}.
Other models of tissues dynamics such as the vertex model~\cite{Manning-2016} has also been considered to take into account of the confluency of biological tissues.
However, recent studies~\cite{Henkes-2020-tissue} showed that there is no significant statistical difference between the vertex model and the particle model 
such as the one we use in this paper.
The particles are confined inside a $2L\times2L$ box with periodic boundary conditions in the $x$-direction and asymmetric hard walls at the top and the bottom 
(see Fig.~\ref{fig:geometry}).
The corrugated walls have periodicity $L$, peak height $h$ and skewness value $\ell\in[0,L]$, see Fig.~\ref{fig:geometry}.
Obviously, $2(L-h)$ has to be larger than the particles' diameter.

% dynamics
We define $\mathbf{r}_i(t)=(x_i(t),y_i(t))$ to be the centre-of-mass position of particle $i$, with $i=1,2,...,N(t)$, where $N(t)$ is the total number of particles. 
Note that $N(t)$ is not conserved since the particles can divide or die. 
We take all $N(t)$ particles to have the same diameter $d$. 
The equation of motion for each particle $i$ is given by the overdamped dynamics:
\begin{equation}
\zeta \frac{d\mathbf{r}_i}{dt}= \sum_{j\neq i} \mathbf{F}_{ij} + \sum_{walls} \mathbf{F}_{iw}, \label{eq:dynamics}
\end{equation}
where $\zeta$ is the friction coefficient,
$\mathbf{F}_{ij}$ is the force exerted on particle $i$ by the neighbouring particle $j$, and
$\mathbf{F}_{iw}$ is the force exerted on particle $i$ by the walls. 
In this model we neglect the inertia and thermal fluctuations of the particles, 
which are justified since the typical Reynolds number of the cells will be of order $10^{-3}$ and cells' diameter will be of order of $10\text{-}100\mu\text{m}$~\cite{Patra-2013-drug-delivery}.
We solve Eq.~(\ref{eq:dynamics}) numerically using Euler update in time with timestep $\Delta t=10^{-3}$.

% particle-particle interaction
For particle-particle interactions, a soft elastic repulsive potential is used, such that the force acting on particle $i$ due to the neighbouring particle $j$ is given by:
\begin{equation} \label{forceij}
  \mathbf{F}_{ij} = kd\left ( 1-\frac{r_{ij}}{d} \right ) H(d-r_{ij}) \hat{\mathbf{r}}_{ij},
\end{equation}
where $k$ is the stiffness, $r_{ij}=|\mathbf{r}_i - \mathbf{r}_j|$ is the distance between particle $i$ and particle $j$ and 
$H(x)$ is the Heaviside step function such that $H(x)=1$ if $x\ge0$ and $H(x)=0$ if $x<0$.
$\hat{\mathbf{r}}_{ij}$ is the unit vector pointing from $\mathbf{r}_j$ to $\mathbf{r}_i$. 
Note that the particles only exert a repulsive force on each other if they overlap, \emph{i.e.} $r_{ij}<d$.

\begin{figure}
\centering
  \includegraphics[width=1.0\columnwidth]{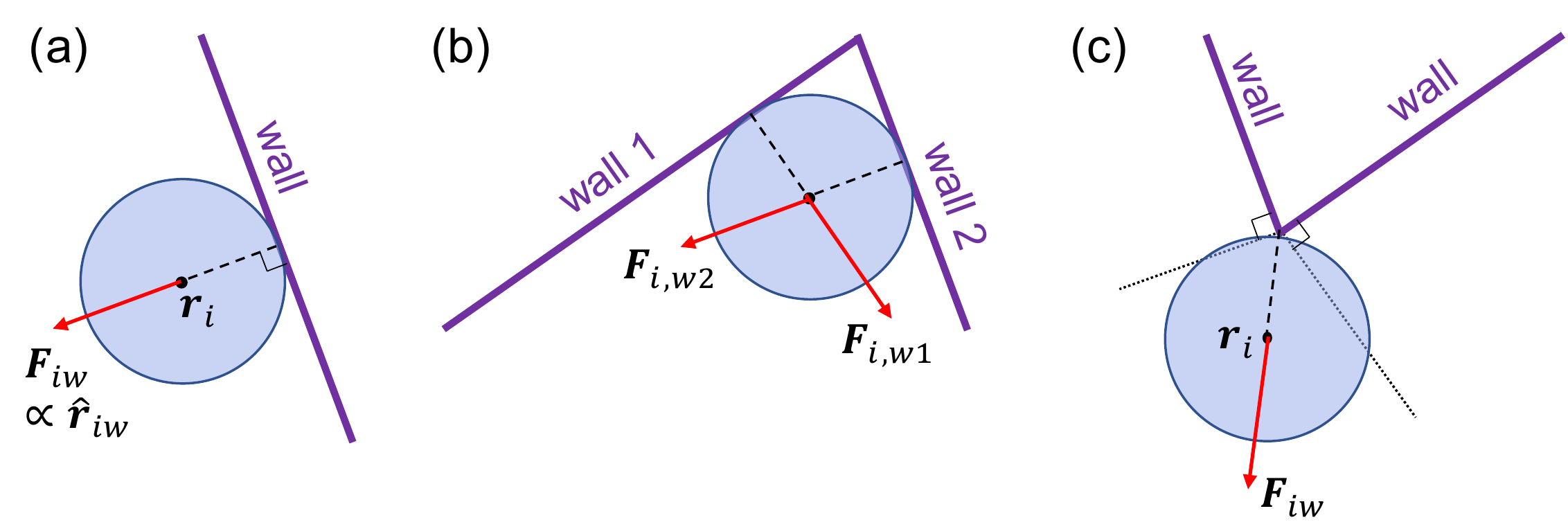}
  \caption{
  Implementations of the wall forces.
  (a) Particle overlaps with one side of the wall: the wall exerts a force $\mathbf{F}_{iw}$ in the perpendicular direction according to Eq.~(\ref{eq:wall}).
  (b) For corner particles, the force from each side of the walls are added up.
  (c) If the particle is in contact with the edge of the wall, the direction of the force points from the edge of the wall to the centre of mass of the particle $\mathbf{r}_i$.
  This only applies if $\mathbf{r}_i$ lies in between the two dotted lines.}
  \label{fig:walls}
\end{figure}

% particle-wall interaction
Now for particle-wall interactions, the same repulsive potential is used, except that the stiffness is made much larger relative to that of particle-particle interactions. 
This is done to approximate the hard walls at the top and bottom of the simulation box.
The resulting force acting on particle $i$ by one side of the walls is then given by: 
\begin{equation}
  \mathbf{F}_{iw} = \frac{\mu kd}{2}\left ( 1-\frac{2r_{iw}}{d} \right ) H\left(\frac{d}{2}-r_{ij}\right) \hat{\mathbf{r}}_{iw}, \label{eq:wall}       
\end{equation}
where $\mu$ is the relative stiffness of the walls (in our simulations, we use $\mu=24$). 
$r_{iw}$ is the perpendicular distance from the centre of mass of the particle $\mathbf{r}_i$ to the wall surface.
$\hat{\mathbf{r}}_{iw}$ is the unit vector pointing from the wall to $\mathbf{r}_i$, in the direction perpendicular to the wall 
[see Fig.~\ref{fig:walls}(a)].
Again the particles only experience repulsive forces from the walls only if they overlap with the walls. 
In the case of a particle in contact with two sides of the walls (corner particles), the wall force from each side of the walls are then added up [see Fig.~\ref{fig:walls}(b)].
Finally for a particle in contact with the edge of the walls [see Fig.~\ref{fig:walls}(c)], 
we first check if the centre of mass of the particle $\mathbf{r}_i$ lies in between the two dotted lines 
[these are two lines which are perpendicular to each respective wall, see Fig.~\ref{fig:walls}(c)]. 
If this is satisfied, the direction of the force is then computed from the edge of the wall to $\mathbf{r}_i$,
otherwise, the rule of Fig.~\ref{fig:walls}(a) applies.

\begin{figure}
\centering
  \includegraphics[width=0.9\columnwidth]{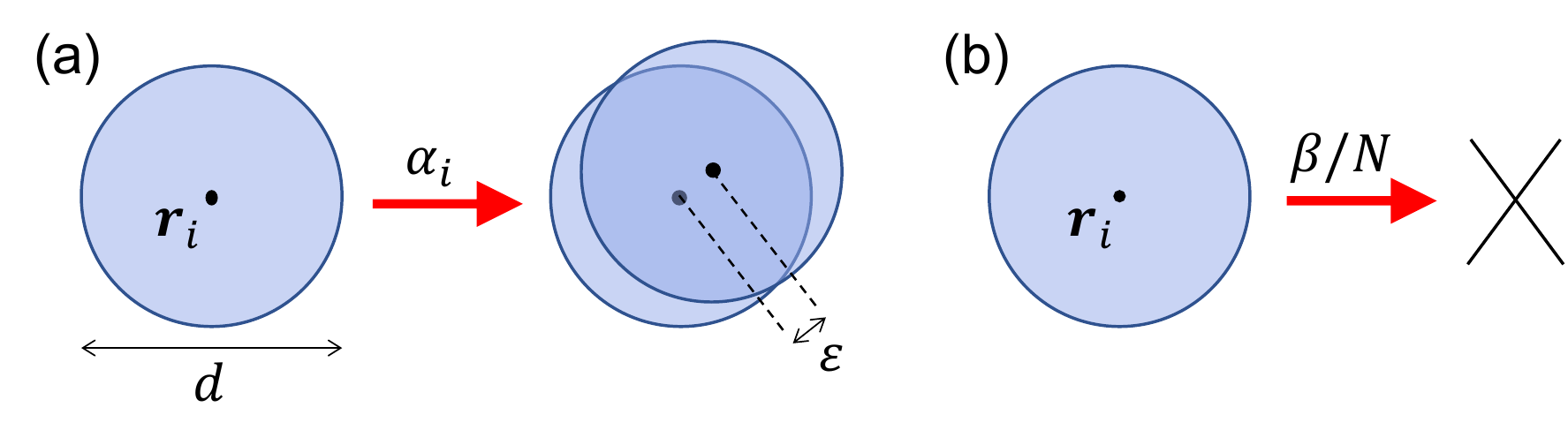}
  \caption{
  (a) Each particle $i$ divides with rate $\alpha_i$, which depends on the number of contacts particle $i$ makes with its neighbour, see Eq.~(\ref{eq:alpha}).
  When the particle divides, a new particle is created at a distance $\varepsilon\ll d$ from the parent particle with random orientation.
  (b) Each particle $i$ also dies with rate $\beta/N$. When the particle dies, it is removed from the system.}
  \label{fig:division}
\end{figure}

% division & death
We also implement particles' division and death as follows.
First, each particle $i$ divides with rate $\alpha_i$, which is defined to be\cite{Barrat-2017-tissue}:
\begin{equation}
\alpha_i = \alpha_0\left(1 - \frac{z_i}{z_{\text{max}}} \right) H(z_{\text{max}}-z_i), \label{eq:alpha}
\end{equation}
where $\alpha_0$ is the rate constant and $z_i$ is the contact number, \emph{i.e.} number of contacts particle $i$ makes with its neighbours.
Overlap with another particle is counted as $1$ contact while overlap with one side of the walls is counted as $2$ contacts.
Overlap with the wall edge [such as one in Fig.~\ref{fig:walls}(c)] is counted as $1$ contact.
Note that the division rate goes to zero when the contact number is larger than or equal to $z_{\text{max}}$.
In this work, we fix $z_{\text{max}}=6$, which corresponds to the hexagonal packing/maximal packing in two-dimensions. 
When particle $i$ divides, we simply add a new particle at a distance $\epsilon\ll d$ from the parent particle in a random direction [see Fig.~\ref{fig:division}(a)].
(In our simulations, we set $\epsilon=0.05d$.)
The daughter and the parent particles then quickly separate out in the subsequent timesteps due to strong repulsion between them. 
Finally, each particle $i$ also dies with rate $\beta/N(t)$. When the particle dies, we simply remove it from the system.
Note that $\beta$ is the \emph{global} death rate, which we fix to be constant in time, whereas $\beta/N(t)$ is the death rate for each particle.
Since we only consider the steady state statistics in this paper, there is no distinction between fixing the global $\beta$ or the individual particle's $\beta/N(t)$ 
to be constant in time.

% units
We shall use the following units for length $[x]=d$ and time $[t]=\zeta/k$.
In these units, the particles' diameter is always $1$ in simulation units. 
$\zeta/k$ is the typical timescale for two particles (separated by a small distance $\ll d$) to separate.
In our simulation units this timescale is again equal to $1$.
All the results below will be presented in simulation units.

\section{Results}

\begin{figure}
\centering
  \includegraphics[width=1.0\columnwidth]{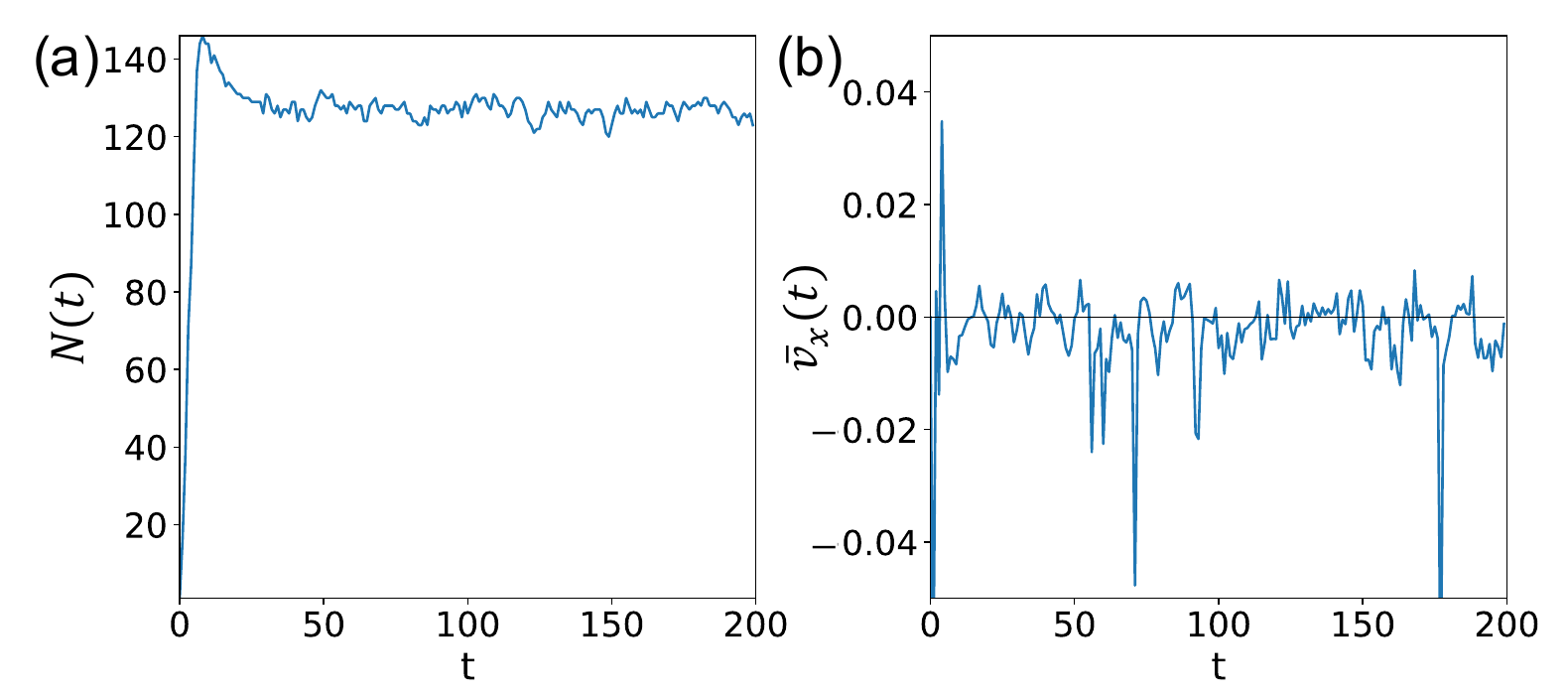}
  \caption{
  (a) Total number of particles $N(t)$ as a function of time $t$ in simulation units for skewness value $\ell/L=0.1$.
  The system reaches a steady state at around $t\simeq50$.
  (b) Average velocity of the particles $\bar{v}_x(t)$ in the $x$-direction as a function of time $t$ for the same run, 
  showing a clear bias towards the negative direction.
  (Parameters used: $L=5$, $h=4.4$, $\alpha_0=20$ and $\beta=2$.)}
  \label{fig:time-evolution}
\end{figure}

\begin{figure}
\centering
  \includegraphics[width=1.0\columnwidth]{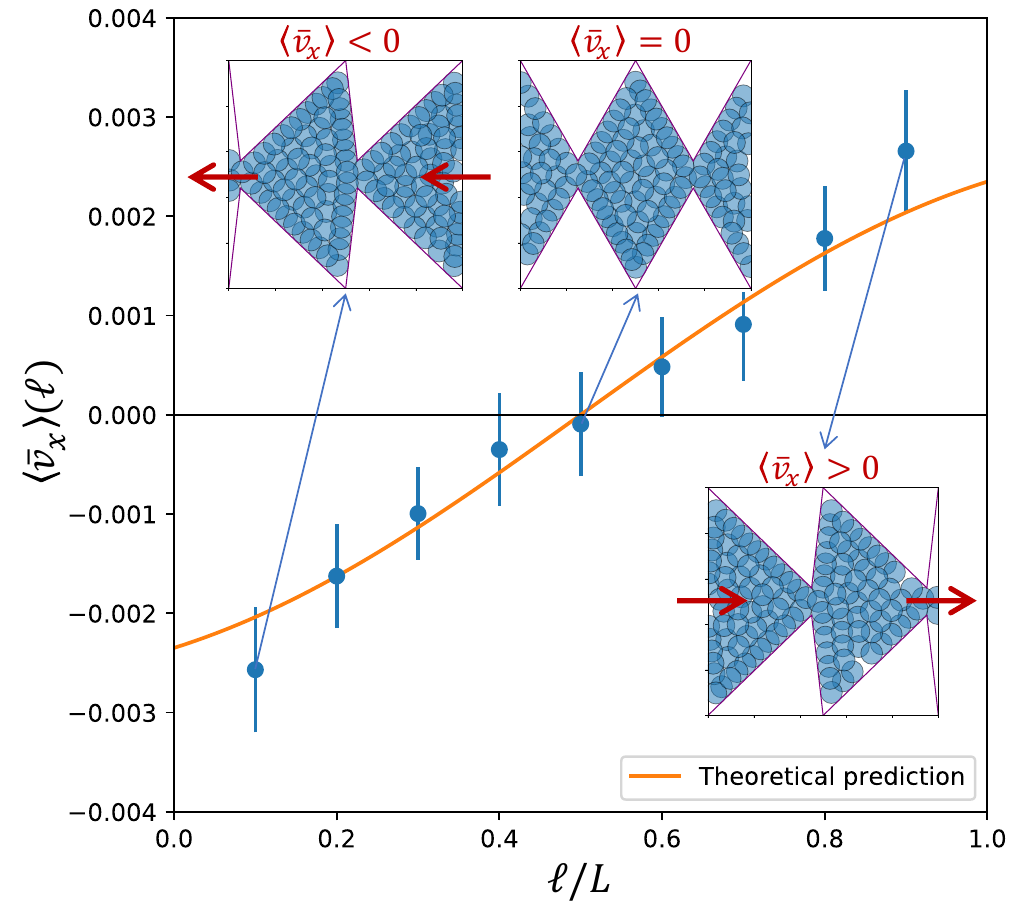}
  \caption{
  Average velocity of the particles in the $x$-direction as a function of the skewness $\ell$.
  Vertical bar on each data point indicates standard deviation. 
  Orange line is the theoretical prediction from Eq.~(\ref{eq:vx-theory}).
  Insets show steady state snapshots for $\ell/L=0.1,0.5,$ and $0.9$.
  (Parameters used: $L=5$, $h=4.4$, $\alpha_0=20$ and $\beta=2$.)
  \label{fig:vx}}
\end{figure}

% N(t)
We initialize all simulations from a single particle located at the centre of the box.
We then let the system multiply and fill the entire box until a steady state is reached (see Supplementary Movies 1).
Fig.~\ref{fig:time-evolution}(a) shows the total number of particles $N(t)$ as a function of time $t$ for one simulation run.
Initially the total number of particles is one: $N(0)=1$, and then $N(t)$ grows exponentially until the steady state is reached at around $t\simeq50$.
At steady state, the total number of particles fluctuate around some average value.
Some snapshots of the steady state configurations are shown in the insets of Fig.~\ref{fig:vx}.
At steady state each particle has an average contact number roughly equal to or more than $z_{\text{max}}=6$, which corresponds to the maximal packing in two-dimension.
At this packing, the particle does not divide anymore since the division rate $\alpha_i$ is zero.
Division is thus induced only if the particle dies (with finite death rate $\beta/N(t)$), at which point,
the contact number drops below $z_{\text{max}}$ and the division rate becomes non-zero again.

% v_x
We also measure the average velocity of the particles in the $x$-direction:
\begin{equation}
\bar{v}_x(t) = \frac{1}{N(t)}\sum_{i=1}^{N(t)}\frac{dx_i}{dt}. 
\end{equation}
The average velocity $\bar{v}_x(t)$ as a function of time $t$ for one simulation run with skewness value $\ell/L=0.1$ is shown in Fig.~\ref{fig:time-evolution}(b).
It shows a clear bias of net velocity in the negative direction, indicating a macroscopic current induced by a combination of local division and apoptosis,
and the asymmetric shape of the channel.

% <v_x>
To make further analysis, we then take the time average and ensemble average of $\bar{v}_x(t)$:
\begin{equation}
\left< \bar{v}_x \right> = \left< \frac{1}{\tau} \int_{t_{\text{ss}}}^{t_\text{ss}+\tau}\bar{v}_x(t)\,dt \right>_e, \label{eq:vx}
\end{equation}
where $t_{\text{ss}}$ is the time it takes for the system to reach the steady state (in our simulations $t_{\text{ss}}\sim50$).
$\tau>0$ is the time length over which we take measurements (in our simulations, we take $\tau=3t_{\text{ss}}$). 
The angle bracket $\left<\dots\right>_e$ in Eq.~(\ref{eq:vx}) indicates ensemble averaging over different simulation runs with the same set of parameter values.
Fig.~\ref{fig:vx} shows the plot of $\left<\bar{v}_x\right>$ as a function of skewness $\ell$.
Each data point is taken from an ensemble average of $100$ independent simulation runs.
In the symmetric case $\ell=0.5L$, the average $x$-velocity is zero as expected.
For $\ell<0.5L$, we observe a net current in the negative $x$-direction, and for $\ell>0.5L$,
we observe a net current in the positive $x$-direction.

% spontaneous flow transition
It is worth noting that in the case of active extensile and contractile fluids such as bacterial suspensions and actomyosin network inside the cells~\cite{Marchetti-2013-review},
a macroscopic current can be observed even along a symmetric channel.
In these examples, the particles are actually elongated and behave as `force dipoles', which act on the surrounding fluid.
A spontaneous symmetry breaking in the orientations of the particles can give rise to a macroscopic flow inside a symmetric channel in either direction
(spontaneous flow transition~\cite{Voituriez-2005,Markovich-2019-polar,Loisy-2018-rheology-polar,Yeomans-2009-spontaneous-flow}).
In our case the particles are spherical and non-swimmers, thus the particles do not have any orientation
(except when they divide into a dumbbell momentarily~\cite{Yeomans-2017-tissue}).
Thus we do not expect to observe any spontaneous flow transition in the symmetric channel \emph{i.e.} $\ell=0.5L$.

\begin{figure}
\centering
  \includegraphics[width=0.6\columnwidth]{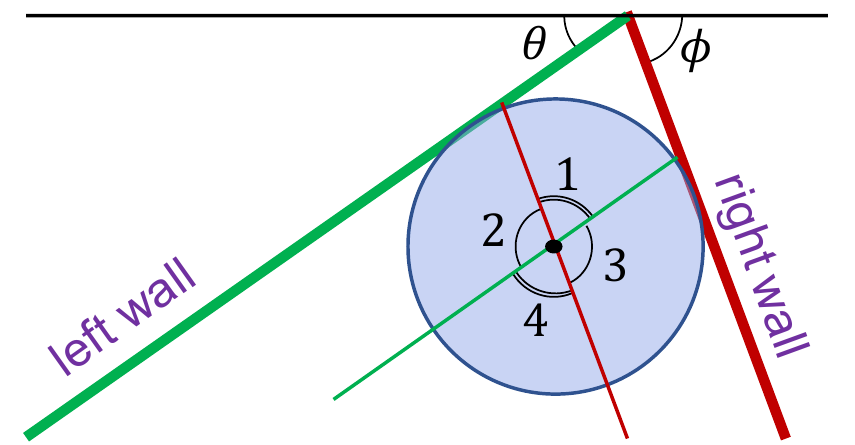}
  \caption{
  Net current of particles in the $x$-direction is mainly caused by division of the corner particles.
  By calculating the wall forces on the daughter particle, the average velocity $\left<\bar{v}_x\right>(\ell)$ as a function of $\ell$ can be computed.
  Green lines are parallel to each other, likewise red lines are parallel to each other.}
  \label{fig:theory}
\end{figure}

% theory
Now we will argue that the net current of particles in the $x$-direction, which we observe in Fig.~\ref{fig:vx},
is mainly caused by particles' division at the corner walls.
Let's consider the geometry of a corner particle in Fig.~\ref{fig:theory}.
Here, $\theta$ and $\phi$ are the slopes of the two adjacent walls.
From Fig.~\ref{fig:geometry}, we can compute: 
\begin{align}
\theta &= \sin^{-1}\left( \frac{h}{\sqrt{(L-\ell)^2+h^2}} \right) \label{eq:theta} \\
\phi &= \sin^{-1}\left( \frac{h}{\sqrt{\ell^2+h^2}} \right) \label{eq:phi}
\end{align}
Let's divide the region around the corner particle into four sectors: $\gamma=1,2,3,$ and $4$ as shown in Fig.~\ref{fig:theory}.
Now suppose that this corner particle divides in the next time step.
Then the probability that we place the centre of mass of the daughter particle inside sector $\gamma$ is proportional to the angle of that sector $\gamma$.
In particular, the probability that the daughter particle is placed inside $\gamma=1$ or $\gamma=4$ is $(\pi-\theta-\phi)/2\pi$
and the probability that the daughter particle is placed inside $\gamma=2$ or $\gamma=3$ is $(\theta+\phi)/2\pi$.

Now suppose that the daughter particle is placed inside sector $\gamma=4$. 
Then the newly created particle will not experience any force from the walls.
However if the daughter particle is placed inside sector $\gamma=2$,
the particle will then experience a wall force from the left wall.
The $x$-component of this force is $F_w\sin\theta$, where $F_w$ is the magnitude of the wall force.
Similarly, if the daughter particle is placed inside sector $\gamma=3$,
the $x$-component of the force coming from the right wall is $-F_w\sin\phi$.
Finally, if the daughter particle is placed inside sector $\gamma=1$, 
the particle will then experience forces from both the left and the right wall.
The $x$-component of this force is: $F_w(\sin\theta-\sin\phi)$.

Therefore, on average, for every division of the corner particle, the walls will exert a force in the $x$-direction with magnitude:
\begin{align}
F_{x,w} &= F_w\left(\frac{\theta+\phi}{2\pi}\sin\theta - \frac{\theta+\phi}{2\pi}\sin\phi +\frac{\pi-\theta-\phi}{2\pi}(\sin\theta-\sin\phi)  \right)  \\
&= \frac{F_w}{2} \left( \frac{h}{\sqrt{(L-\ell)^2+h^2}} - \frac{h}{\sqrt{\ell^2+h^2}} \right),
\end{align}
where we have substituted Eqns.~(\ref{eq:theta}-\ref{eq:phi}) in the last line above.
Now we can assume that the average velocity of all the particles is proportional to the wall forces exerted on the system: $\left<\bar{v}_x\right>\propto F_{x,w}$
(in other words, the tissue responds like a Newtonian fluid).
We can then estimate the average velocity in the $x$-direction as a function of the skewness $\ell$:
\begin{equation}
\left<\bar{v}_x\right>(\ell) \propto \left( \frac{h}{\sqrt{(L-\ell)^2+h^2}} - \frac{h}{\sqrt{\ell^2+h^2}} \right). \label{eq:vx-theory}
\end{equation}
Fig.~\ref{fig:vx} shows the comparison between the theoretical estimate (orange line) with the simulation data.
In Fig.~\ref{fig:vx}, the proportionality constant in Eq.~(\ref{eq:vx-theory}) is obtained from the best fit value.

% discrepancy
Although the simple theoretical model in Eq.~(\ref{eq:vx-theory}) captures the direction of the current correctly,
the curvature of $\left<\bar{v}_x\right>(\ell)$ does not fit perfectly to the simulation data, especially at extreme skewness values: $\ell\simeq0.1L$ and $\ell\simeq0.9L$.
This can be due to several factors.
Firstly, the motions of the particles at high density are highly correlated.
The force due to particles' divisions in the bulk can propagate to the walls~\cite{Barrat-2018-elastoplastic}.
Secondly, from Fig.~\ref{fig:time-evolution}(b), we also see occasional spikes of rather large negative net velocity. 
This is due to sudden jamming and unjamming of particles near the narrow gap between the top and the bottom wall, 
which were not accounted for in the theory Eq.~(\ref{eq:vx-theory}).

\begin{figure}
\centering
  \includegraphics[width=1.0\columnwidth]{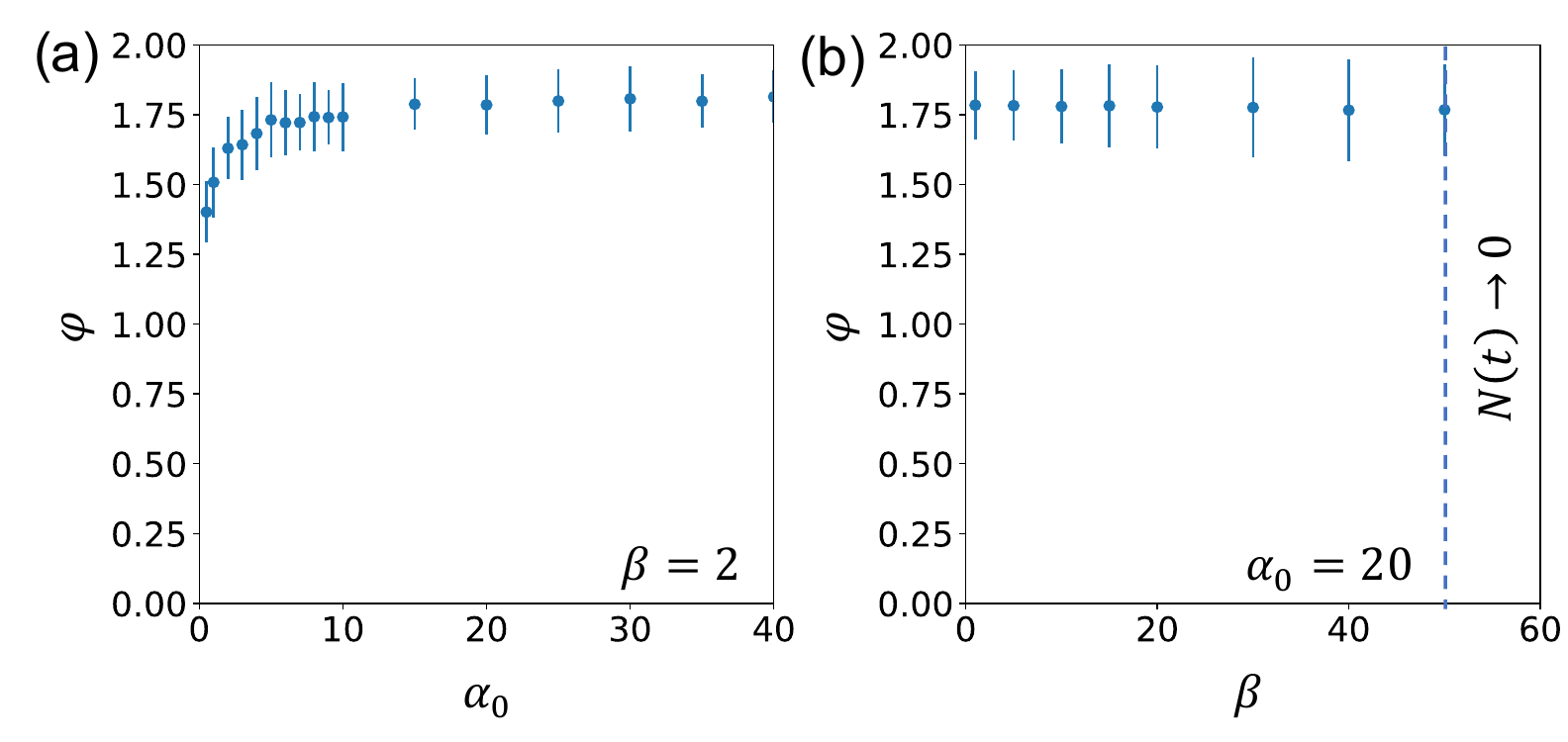}
  \caption{
  (a) The steady state area fraction $\varphi$ as a function of division rate $\alpha_0$ for fixed global death rate $\beta=2$.
  (b) The steady state area fraction $\varphi$ as a function of global death rate $\beta$ for a fixed division rate $\alpha_0=20$.
  (Parameters used: $L=5$, $h=4.4$, and $\ell=0.5L$.)}
  \label{fig:phi}
\end{figure}

% area fraction
We also measured the area fraction of the system in the steady state, which is defined to be the total area of the particles divided by the area available for the particles:
\begin{equation}
\varphi = \frac{\left<N\right>\pi d^2}{4(4L^2 - 2Lh)},
\end{equation}
where $d$ is the particles' diameter and $\left<N\right>$ is the average number of particles in the steady state.
Fig.~\ref{fig:phi}(a) shows the steady state area fraction $\varphi$ as a function of division rate $\alpha_0$ for fixed global death rate $\beta=2$.
As we can see from the figure, $\varphi$ increases then saturates with increasing division rate as expected.
Fig.~\ref{fig:phi}(b) shows the steady state area fraction $\varphi$ as a function of global death rate $\beta$ for a fixed division rate $\alpha_0=20$.
Here, $\varphi$ does not actually vary much with $\beta$, except for a very large value of $\beta\gtrsim50$ [dashed line in Fig.~\ref{fig:phi}(b)].
In this regime $\beta\gtrsim50$, the death rate is so fast compared to $\alpha_0$ so that the number of particles goes to zero.

From Fig.~\ref{fig:phi}, we also note that the area fraction in the steady state appears to be larger than one: $\varphi\simeq1.8$ 
for a large range of parameters.
This is because a soft spring potential is used to model the interactions between the particles.
This potential allows the particles to overlap, which mimic the mechanics of deformable particles such as foams~\cite{Durian-1995}.
Thus the diameter $d$ in our model does not necessarily correspond to the diameter of the real cells (which are also deformable).
The total overlaps between the particles also give rise to an internal pressure, which corresponds to the homeostatic pressure, 
observed in biological tissues~\cite{Prost-2009,Montel-2011}.

\begin{figure}
\centering
  \includegraphics[width=1.0\columnwidth]{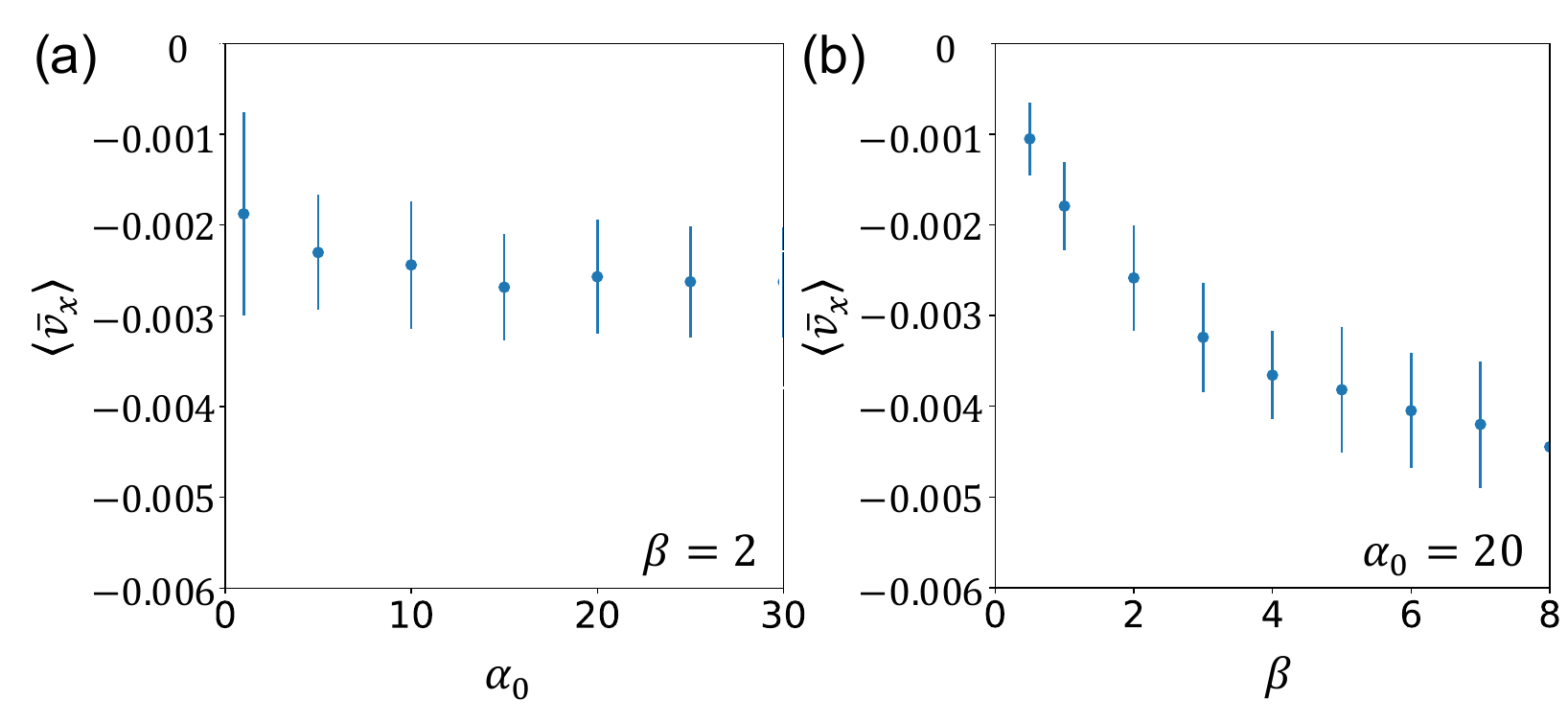}
  \caption{
  (a) Average velocity as a function of division rate $\alpha_0$ for fixed $\beta=2$.
  (b) Average velocity as a function of global death rate $\beta$ for fixed $\alpha_0=20$.
  (Parameters used: $L=5$, $h=4.4$, and $\ell=0.1L$.)}
  \label{fig:vx-alpha-beta}
\end{figure}

% velocity dependence on alpha and beta
Finally Fig.~\ref{fig:vx-alpha-beta} shows the dependence of the average $x$-velocity $\left<\bar{v}_x\right>$ on the division rate $\alpha_0$ and the global death rate $\beta$.
As we can see from the figure, $\left<\bar{v}_x\right>$ is affected more strongly by the death rate $\beta/\left<N\right>$ than the division rate $\alpha_0$.
As we argue in Fig.~\ref{fig:theory}, the average velocity is mostly controlled by the division of the corner particles.
At steady state, the corner particles have an average contact number close to $z_{\text{max}}=6$.
Accordingly, the division rate of the particles is close to zero, see Eq.~(\ref{eq:alpha}).
Division is only triggered when one of the neighbouring particles dies with rate $\beta/N(t)$, which then reduces the number of contacts.
Therefore $\beta$ sets the limiting rate for the division of the corner particles.

\section{Conclusions}
We have shown that local division and apoptosis events are sufficient to give rise to a macroscopic current (or work),
without any self-propulsion mechanism.
We have found that the maximum average velocity, obtained in our ratchet geometry is around $0.003$ in simulation units.
To convert this to physical units, we note that the typical cell size is around $d\simeq100\,\mu\text{m}$ and 
the apoptosis rate is around $\beta/\left<N\right>\simeq10^{-6}\,\text{s}^{-1}$~\cite{Thirumalai-2018}.
We can then set the units of length and time in our simulations to be $100\,\mu\text{m}$ and $\text{hour}$ respectively.
This translates to a maximum velocity of around $0.3\,\mu\text{m}/\text{hour}$, which is significantly smaller than that of bacterial ratchet.
Thus although local division and apoptosis are sufficient to give macroscopic work, they are much less efficient than rectified self-propelled particles.

\begin{figure}
\centering
  \includegraphics[width=0.6\columnwidth]{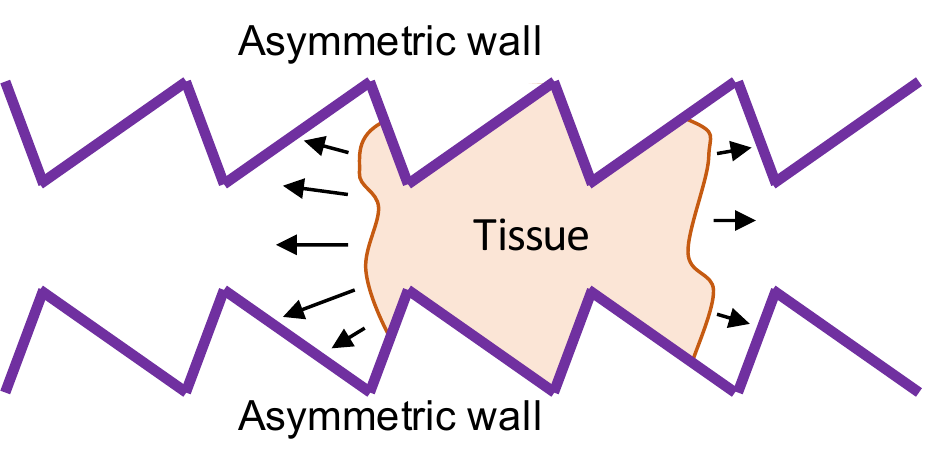}
  \caption{
  A growing tissue (or bacterial colony, \emph{etc.}) might grow asymmetrically due to the ratchet effect.}
  \label{fig:experiment}
\end{figure}

One way to improve the efficiency of our computational model is to consider the growth phase of the tissue rather than the steady state behaviour.
In this paper we use a periodic boundary condition in the $x$-direction and the measurements are always taken in the steady state.
However, at steady state, the division rate of the corner particles (which are responsible for the net current) is limited by the death rate $\beta/\left<N\right>$.
On the other hand, when we consider the growth phase of the tissue, where the total number of particles $N(t)$ is still growing exponentially, 
the division rates of the corner particles are much faster and therefore we expect the net velocity to be much higher.
To test this, we need to perform simulations in a sufficiently long channel without periodic boundary conditions and let the tissue grow into the empty space of this extremely long channel (see Fig.~\ref{fig:experiment}).
We then expect the tissue to grow asymmetrically due to the ratchet effect, explained in this paper.

It might be interesting to replicate our ratchet geometry in real biological tissues (or growing bacterial colonies) to detect any net current such as the one depicted in Fig.~\ref{fig:experiment}.
However, some cells such as epithelial cells are also motile~\cite{Angelini-2011} (in addition to local division and apoptosis events),
which makes it difficult to disentangle the physics of local division and apoptosis from self-propulsion.
Other type of cells such as most tumour cells~\cite{Poincloux-2011} and biofilms~\cite{DellArciprete-2018}, for example, are much less motile and may be more suitable to test our theory. 

Finally, our results can have potential applications in tissue engineering, for example, in treating achilles tendon injuries.
One might engineer a ratchet-shaped scaffold starting from either end of the injured tendon, directed towards the middle, to speed up the healing process.

\section*{Author Contributions}
Both authors contributed equally to this work.

\section*{Conflicts of interest}
There are no conflicts to declare.

\section*{Acknowledgements}
The authors acknowledge useful discussion with Matteo Degiacomi and Suzanne Fielding.

%%%REFERENCES%%%
\bibliography{rsc} %You need to replace "rsc" on this line with the name of your .bib file

\begin{thebibliography}{25}%
\makeatletter
\providecommand \@ifxundefined [1]{%
 \@ifx{#1\undefined}
}%
\providecommand \@ifnum [1]{%
 \ifnum #1\expandafter \@firstoftwo
 \else \expandafter \@secondoftwo
 \fi
}%
\providecommand \@ifx [1]{%
 \ifx #1\expandafter \@firstoftwo
 \else \expandafter \@secondoftwo
 \fi
}%
\providecommand \natexlab [1]{#1}%
\providecommand \enquote  [1]{``#1''}%
\providecommand \bibnamefont  [1]{#1}%
\providecommand \bibfnamefont [1]{#1}%
\providecommand \citenamefont [1]{#1}%
\providecommand \href@noop [0]{\@secondoftwo}%
\providecommand \href [0]{\begingroup \@sanitize@url \@href}%
\providecommand \@href[1]{\@@startlink{#1}\@@href}%
\providecommand \@@href[1]{\endgroup#1\@@endlink}%
\providecommand \@sanitize@url [0]{\catcode `\\12\catcode `\$12\catcode
  `\&12\catcode `\#12\catcode `\^12\catcode `\_12\catcode `\%12\relax}%
\providecommand \@@startlink[1]{}%
\providecommand \@@endlink[0]{}%
\providecommand \url  [0]{\begingroup\@sanitize@url \@url }%
\providecommand \@url [1]{\endgroup\@href {#1}{\urlprefix }}%
\providecommand \urlprefix  [0]{URL }%
\providecommand \Eprint [0]{\href }%
\providecommand \doibase [0]{http://dx.doi.org/}%
\providecommand \selectlanguage [0]{\@gobble}%
\providecommand \bibinfo  [0]{\@secondoftwo}%
\providecommand \bibfield  [0]{\@secondoftwo}%
\providecommand \translation [1]{[#1]}%
\providecommand \BibitemOpen [0]{}%
\providecommand \bibitemStop [0]{}%
\providecommand \bibitemNoStop [0]{.\EOS\space}%
\providecommand \EOS [0]{\spacefactor3000\relax}%
\providecommand \BibitemShut  [1]{\csname bibitem#1\endcsname}%
\let\auto@bib@innerbib\@empty
%</preamble>
\bibitem [{\citenamefont {Di~Leonardo}\ \emph {et~al.}(2010)\citenamefont
  {Di~Leonardo}, \citenamefont {Angelani}, \citenamefont
  {Dell{\textquoteright}Arciprete}, \citenamefont {Ruocco}, \citenamefont
  {Iebba}, \citenamefont {Schippa}, \citenamefont {Conte}, \citenamefont
  {Mecarini}, \citenamefont {De~Angelis},\ and\ \citenamefont
  {Di~Fabrizio}}]{DiLeonardo-2010-ratchet}%
  \BibitemOpen
  \bibfield  {author} {\bibinfo {author} {\bibfnamefont {R.}~\bibnamefont
  {Di~Leonardo}}, \bibinfo {author} {\bibfnamefont {L.}~\bibnamefont
  {Angelani}}, \bibinfo {author} {\bibfnamefont {D.}~\bibnamefont
  {Dell{\textquoteright}Arciprete}}, \bibinfo {author} {\bibfnamefont
  {G.}~\bibnamefont {Ruocco}}, \bibinfo {author} {\bibfnamefont
  {V.}~\bibnamefont {Iebba}}, \bibinfo {author} {\bibfnamefont
  {S.}~\bibnamefont {Schippa}}, \bibinfo {author} {\bibfnamefont {M.~P.}\
  \bibnamefont {Conte}}, \bibinfo {author} {\bibfnamefont {F.}~\bibnamefont
  {Mecarini}}, \bibinfo {author} {\bibfnamefont {F.}~\bibnamefont
  {De~Angelis}}, \ and\ \bibinfo {author} {\bibfnamefont {E.}~\bibnamefont
  {Di~Fabrizio}},\ }\href {\doibase 10.1073/pnas.0910426107} {\bibfield
  {journal} {\bibinfo  {journal} {Proceedings of the National Academy of
  Sciences}\ }\textbf {\bibinfo {volume} {107}},\ \bibinfo {pages} {9541}
  (\bibinfo {year} {2010})},\ \Eprint
  {http://arxiv.org/abs/https://www.pnas.org/content/107/21/9541.full.pdf}
  {https://www.pnas.org/content/107/21/9541.full.pdf} \BibitemShut {NoStop}%
\bibitem [{\citenamefont {Galajda}\ \emph {et~al.}(2007)\citenamefont
  {Galajda}, \citenamefont {Keymer}, \citenamefont {Chaikin},\ and\
  \citenamefont {Austin}}]{Galajda-2007-ratchet}%
  \BibitemOpen
  \bibfield  {author} {\bibinfo {author} {\bibfnamefont {P.}~\bibnamefont
  {Galajda}}, \bibinfo {author} {\bibfnamefont {J.}~\bibnamefont {Keymer}},
  \bibinfo {author} {\bibfnamefont {P.}~\bibnamefont {Chaikin}}, \ and\
  \bibinfo {author} {\bibfnamefont {R.}~\bibnamefont {Austin}},\ }\href
  {\doibase 10.1128/JB.01033-07} {\bibfield  {journal} {\bibinfo  {journal} {J.
  Bacteriol.}\ }\textbf {\bibinfo {volume} {189}},\ \bibinfo {pages} {8704}
  (\bibinfo {year} {2007})}\BibitemShut {NoStop}%
\bibitem [{\citenamefont {Woillez}\ \emph {et~al.}(2020)\citenamefont
  {Woillez}, \citenamefont {Kafri},\ and\ \citenamefont
  {Lecomte}}]{Lecomte-2020-ratchet}%
  \BibitemOpen
  \bibfield  {author} {\bibinfo {author} {\bibfnamefont {E.}~\bibnamefont
  {Woillez}}, \bibinfo {author} {\bibfnamefont {Y.}~\bibnamefont {Kafri}}, \
  and\ \bibinfo {author} {\bibfnamefont {V.}~\bibnamefont {Lecomte}},\ }\href
  {\doibase https://doi.org/10.1088/1742-5468/ab7e2e} {\bibfield  {journal}
  {\bibinfo  {journal} {J. Stat. Mech.}\ }\textbf {\bibinfo {volume}
  {063204}},\ \bibinfo {pages} {1} (\bibinfo {year} {2020})}\BibitemShut
  {NoStop}%
\bibitem [{\citenamefont {Stenhammar}\ \emph {et~al.}(2016)\citenamefont
  {Stenhammar}, \citenamefont {Wittkowski}, \citenamefont {Marenduzzo},\ and\
  \citenamefont {Cates}}]{Stenhammar-2016-ratchet}%
  \BibitemOpen
  \bibfield  {author} {\bibinfo {author} {\bibfnamefont {J.}~\bibnamefont
  {Stenhammar}}, \bibinfo {author} {\bibfnamefont {R.}~\bibnamefont
  {Wittkowski}}, \bibinfo {author} {\bibfnamefont {D.}~\bibnamefont
  {Marenduzzo}}, \ and\ \bibinfo {author} {\bibfnamefont {M.~E.}\ \bibnamefont
  {Cates}},\ }\href {\doibase 10.1126/sciadv.1501850} {\bibfield  {journal}
  {\bibinfo  {journal} {Science Advances}\ }\textbf {\bibinfo {volume} {2}}
  (\bibinfo {year} {2016}),\ 10.1126/sciadv.1501850},\ \Eprint
  {http://arxiv.org/abs/https://advances.sciencemag.org/content/2/4/e1501850.full.pdf}
  {https://advances.sciencemag.org/content/2/4/e1501850.full.pdf} \BibitemShut
  {NoStop}%
\bibitem [{\citenamefont {Reichhardt}\ and\ \citenamefont
  {Reichhardt}(2017)}]{Reichhardt-2017-ratchet}%
  \BibitemOpen
  \bibfield  {author} {\bibinfo {author} {\bibfnamefont {C.~J.~O.}\
  \bibnamefont {Reichhardt}}\ and\ \bibinfo {author} {\bibfnamefont
  {C.}~\bibnamefont {Reichhardt}},\ }\href {\doibase
  10.1146/annurev-conmatphys-031016-025522} {\bibfield  {journal} {\bibinfo
  {journal} {Annual Review of Condensed Matter Physics}\ }\textbf {\bibinfo
  {volume} {8}},\ \bibinfo {pages} {51} (\bibinfo {year} {2017})},\ \Eprint
  {http://arxiv.org/abs/https://doi.org/10.1146/annurev-conmatphys-031016-025522}
  {https://doi.org/10.1146/annurev-conmatphys-031016-025522} \BibitemShut
  {NoStop}%
\bibitem [{\citenamefont {Reichhardt}\ and\ \citenamefont
  {Reichhardt}(2013)}]{Reichhardt-2013-ratchet}%
  \BibitemOpen
  \bibfield  {author} {\bibinfo {author} {\bibfnamefont {C.}~\bibnamefont
  {Reichhardt}}\ and\ \bibinfo {author} {\bibfnamefont {C.~J.~O.}\ \bibnamefont
  {Reichhardt}},\ }\href {\doibase 10.1103/PhysRevE.88.062310} {\bibfield
  {journal} {\bibinfo  {journal} {Phys. Rev. E}\ }\textbf {\bibinfo {volume}
  {88}},\ \bibinfo {pages} {062310} (\bibinfo {year} {2013})}\BibitemShut
  {NoStop}%
\bibitem [{\citenamefont {Voituriez}\ \emph {et~al.}(2005)\citenamefont
  {Voituriez}, \citenamefont {Joanny},\ and\ \citenamefont
  {Prost}}]{Voituriez-2005}%
  \BibitemOpen
  \bibfield  {author} {\bibinfo {author} {\bibfnamefont {R.}~\bibnamefont
  {Voituriez}}, \bibinfo {author} {\bibfnamefont {J.~F.-.}\ \bibnamefont
  {Joanny}}, \ and\ \bibinfo {author} {\bibfnamefont {J.}~\bibnamefont
  {Prost}},\ }\href {\doibase 10.1209/epl/i2004-10501-2} {\bibfield  {journal}
  {\bibinfo  {journal} {Europhysics Letters ({EPL})}\ }\textbf {\bibinfo
  {volume} {70}},\ \bibinfo {pages} {404} (\bibinfo {year} {2005})}\BibitemShut
  {NoStop}%
\bibitem [{\citenamefont {Markovich}\ \emph {et~al.}(2019)\citenamefont
  {Markovich}, \citenamefont {Tjhung},\ and\ \citenamefont
  {Cates}}]{Markovich-2019-polar}%
  \BibitemOpen
  \bibfield  {author} {\bibinfo {author} {\bibfnamefont {T.}~\bibnamefont
  {Markovich}}, \bibinfo {author} {\bibfnamefont {E.}~\bibnamefont {Tjhung}}, \
  and\ \bibinfo {author} {\bibfnamefont {M.~E.}\ \bibnamefont {Cates}},\ }\href
  {\doibase 10.1103/PhysRevLett.122.088004} {\bibfield  {journal} {\bibinfo
  {journal} {Phys. Rev. Lett.}\ }\textbf {\bibinfo {volume} {122}},\ \bibinfo
  {pages} {088004} (\bibinfo {year} {2019})}\BibitemShut {NoStop}%
\bibitem [{\citenamefont {Loisy}\ \emph {et~al.}(2018)\citenamefont {Loisy},
  \citenamefont {Eggers},\ and\ \citenamefont
  {Liverpool}}]{Loisy-2018-rheology-polar}%
  \BibitemOpen
  \bibfield  {author} {\bibinfo {author} {\bibfnamefont {A.}~\bibnamefont
  {Loisy}}, \bibinfo {author} {\bibfnamefont {J.}~\bibnamefont {Eggers}}, \
  and\ \bibinfo {author} {\bibfnamefont {T.~B.}\ \bibnamefont {Liverpool}},\
  }\href {\doibase 10.1103/PhysRevLett.121.018001} {\bibfield  {journal}
  {\bibinfo  {journal} {Phys. Rev. Lett.}\ }\textbf {\bibinfo {volume} {121}},\
  \bibinfo {pages} {018001} (\bibinfo {year} {2018})}\BibitemShut {NoStop}%
\bibitem [{\citenamefont {Marchetti}\ \emph {et~al.}(2013)\citenamefont
  {Marchetti}, \citenamefont {Joanny}, \citenamefont {Ramaswamy}, \citenamefont
  {Liverpool}, \citenamefont {Prost}, \citenamefont {Rao},\ and\ \citenamefont
  {Simha}}]{Marchetti-2013-review}%
  \BibitemOpen
  \bibfield  {author} {\bibinfo {author} {\bibfnamefont {M.~C.}\ \bibnamefont
  {Marchetti}}, \bibinfo {author} {\bibfnamefont {J.~F.-.}\ \bibnamefont
  {Joanny}}, \bibinfo {author} {\bibfnamefont {S.}~\bibnamefont {Ramaswamy}},
  \bibinfo {author} {\bibfnamefont {T.~B.}\ \bibnamefont {Liverpool}}, \bibinfo
  {author} {\bibfnamefont {J.}~\bibnamefont {Prost}}, \bibinfo {author}
  {\bibfnamefont {M.}~\bibnamefont {Rao}}, \ and\ \bibinfo {author}
  {\bibfnamefont {R.~A.}\ \bibnamefont {Simha}},\ }\href {\doibase
  10.1103/RevModPhys.85.1143} {\bibfield  {journal} {\bibinfo  {journal} {Rev.
  Mod. Phys.}\ }\textbf {\bibinfo {volume} {85}},\ \bibinfo {pages} {1143}
  (\bibinfo {year} {2013})}\BibitemShut {NoStop}%
\bibitem [{\citenamefont {Matoz-Fernandez}\ \emph {et~al.}(2017)\citenamefont
  {Matoz-Fernandez}, \citenamefont {Agoritsas}, \citenamefont {Barrat},
  \citenamefont {Bertin},\ and\ \citenamefont {Martens}}]{Barrat-2017-tissue}%
  \BibitemOpen
  \bibfield  {author} {\bibinfo {author} {\bibfnamefont {D.~A.}\ \bibnamefont
  {Matoz-Fernandez}}, \bibinfo {author} {\bibfnamefont {E.}~\bibnamefont
  {Agoritsas}}, \bibinfo {author} {\bibfnamefont {J.-L.}\ \bibnamefont
  {Barrat}}, \bibinfo {author} {\bibfnamefont {E.}~\bibnamefont {Bertin}}, \
  and\ \bibinfo {author} {\bibfnamefont {K.}~\bibnamefont {Martens}},\ }\href
  {\doibase 10.1103/PhysRevLett.118.158105} {\bibfield  {journal} {\bibinfo
  {journal} {Phys. Rev. Lett.}\ }\textbf {\bibinfo {volume} {118}},\ \bibinfo
  {pages} {158105} (\bibinfo {year} {2017})}\BibitemShut {NoStop}%
\bibitem [{\citenamefont {Henkes}\ \emph {et~al.}(2020)\citenamefont {Henkes},
  \citenamefont {Kostanjevec}, \citenamefont {Collinson}, \citenamefont
  {Sknepkek},\ and\ \citenamefont {E.}}]{Henkes-2020-tissue}%
  \BibitemOpen
  \bibfield  {author} {\bibinfo {author} {\bibfnamefont {S.}~\bibnamefont
  {Henkes}}, \bibinfo {author} {\bibfnamefont {K.}~\bibnamefont {Kostanjevec}},
  \bibinfo {author} {\bibfnamefont {J.~M.}\ \bibnamefont {Collinson}}, \bibinfo
  {author} {\bibfnamefont {R.}~\bibnamefont {Sknepkek}}, \ and\ \bibinfo
  {author} {\bibfnamefont {B.}~\bibnamefont {E.}},\ }\href {\doibase
  https://doi.org/10.1038/s41467-020-15164-5} {\bibfield  {journal} {\bibinfo
  {journal} {Nat. Comms.}\ }\textbf {\bibinfo {volume} {11}},\ \bibinfo {pages}
  {1} (\bibinfo {year} {2020})}\BibitemShut {NoStop}%
\bibitem [{\citenamefont {Tjhung}\ and\ \citenamefont
  {Berthier}(2020)}]{Tjhung-Berthier-2020}%
  \BibitemOpen
  \bibfield  {author} {\bibinfo {author} {\bibfnamefont {E.}~\bibnamefont
  {Tjhung}}\ and\ \bibinfo {author} {\bibfnamefont {L.}~\bibnamefont
  {Berthier}},\ }\href {\doibase 10.1103/PhysRevResearch.2.043334} {\bibfield
  {journal} {\bibinfo  {journal} {Phys. Rev. Research}\ }\textbf {\bibinfo
  {volume} {2}},\ \bibinfo {pages} {043334} (\bibinfo {year}
  {2020})}\BibitemShut {NoStop}%
\bibitem [{\citenamefont {Bi}\ \emph {et~al.}(2016)\citenamefont {Bi},
  \citenamefont {Yang}, \citenamefont {Marchetti},\ and\ \citenamefont
  {Manning}}]{Manning-2016}%
  \BibitemOpen
  \bibfield  {author} {\bibinfo {author} {\bibfnamefont {D.}~\bibnamefont
  {Bi}}, \bibinfo {author} {\bibfnamefont {X.}~\bibnamefont {Yang}}, \bibinfo
  {author} {\bibfnamefont {M.~C.}\ \bibnamefont {Marchetti}}, \ and\ \bibinfo
  {author} {\bibfnamefont {M.~L.}\ \bibnamefont {Manning}},\ }\href {\doibase
  10.1103/PhysRevX.6.021011} {\bibfield  {journal} {\bibinfo  {journal} {Phys.
  Rev. X}\ }\textbf {\bibinfo {volume} {6}},\ \bibinfo {pages} {021011}
  (\bibinfo {year} {2016})}\BibitemShut {NoStop}%
\bibitem [{\citenamefont {Patra}\ \emph {et~al.}(2013)\citenamefont {Patra},
  \citenamefont {Sengupta}, \citenamefont {Duan}, \citenamefont {Zhang},
  \citenamefont {Pavlick},\ and\ \citenamefont
  {Sen}}]{Patra-2013-drug-delivery}%
  \BibitemOpen
  \bibfield  {author} {\bibinfo {author} {\bibfnamefont {D.}~\bibnamefont
  {Patra}}, \bibinfo {author} {\bibfnamefont {S.}~\bibnamefont {Sengupta}},
  \bibinfo {author} {\bibfnamefont {W.}~\bibnamefont {Duan}}, \bibinfo {author}
  {\bibfnamefont {H.}~\bibnamefont {Zhang}}, \bibinfo {author} {\bibfnamefont
  {R.}~\bibnamefont {Pavlick}}, \ and\ \bibinfo {author} {\bibfnamefont
  {A.}~\bibnamefont {Sen}},\ }\href {\doibase 10.1039/C2NR32600K} {\bibfield
  {journal} {\bibinfo  {journal} {Nanoscale}\ }\textbf {\bibinfo {volume}
  {5}},\ \bibinfo {pages} {1273} (\bibinfo {year} {2013})}\BibitemShut
  {NoStop}%
\bibitem [{\citenamefont {Edwards}\ and\ \citenamefont
  {Yeomans}(2009)}]{Yeomans-2009-spontaneous-flow}%
  \BibitemOpen
  \bibfield  {author} {\bibinfo {author} {\bibfnamefont {S.~A.}\ \bibnamefont
  {Edwards}}\ and\ \bibinfo {author} {\bibfnamefont {J.~M.}\ \bibnamefont
  {Yeomans}},\ }\href {\doibase 10.1209/0295-5075/85/18008} {\bibfield
  {journal} {\bibinfo  {journal} {{EPL} (Europhysics Letters)}\ }\textbf
  {\bibinfo {volume} {85}},\ \bibinfo {pages} {18008} (\bibinfo {year}
  {2009})}\BibitemShut {NoStop}%
\bibitem [{\citenamefont {Saw}\ \emph {et~al.}(2017)\citenamefont {Saw},
  \citenamefont {Doostmohammadi}, \citenamefont {Nier}, \citenamefont
  {Kocgozlu}, \citenamefont {Thampi}, \citenamefont {Toyama}, \citenamefont
  {Marcq}, \citenamefont {Lim}, \citenamefont {Yeomans},\ and\ \citenamefont
  {Ladoux}}]{Yeomans-2017-tissue}%
  \BibitemOpen
  \bibfield  {author} {\bibinfo {author} {\bibfnamefont {T.~B.}\ \bibnamefont
  {Saw}}, \bibinfo {author} {\bibfnamefont {A.}~\bibnamefont {Doostmohammadi}},
  \bibinfo {author} {\bibfnamefont {V.}~\bibnamefont {Nier}}, \bibinfo {author}
  {\bibfnamefont {L.}~\bibnamefont {Kocgozlu}}, \bibinfo {author}
  {\bibfnamefont {S.}~\bibnamefont {Thampi}}, \bibinfo {author} {\bibfnamefont
  {Y.}~\bibnamefont {Toyama}}, \bibinfo {author} {\bibfnamefont
  {P.}~\bibnamefont {Marcq}}, \bibinfo {author} {\bibfnamefont {C.~T.}\
  \bibnamefont {Lim}}, \bibinfo {author} {\bibfnamefont {J.~M.}\ \bibnamefont
  {Yeomans}}, \ and\ \bibinfo {author} {\bibfnamefont {B.}~\bibnamefont
  {Ladoux}},\ }\href@noop {} {\bibfield  {journal} {\bibinfo  {journal}
  {Nature}\ }\textbf {\bibinfo {volume} {544}},\ \bibinfo {pages} {212}
  (\bibinfo {year} {2017})}\BibitemShut {NoStop}%
\bibitem [{\citenamefont {Nicolas}\ \emph {et~al.}(2018)\citenamefont
  {Nicolas}, \citenamefont {Ferrero}, \citenamefont {Martens},\ and\
  \citenamefont {Barrat}}]{Barrat-2018-elastoplastic}%
  \BibitemOpen
  \bibfield  {author} {\bibinfo {author} {\bibfnamefont {A.}~\bibnamefont
  {Nicolas}}, \bibinfo {author} {\bibfnamefont {E.~E.}\ \bibnamefont
  {Ferrero}}, \bibinfo {author} {\bibfnamefont {K.}~\bibnamefont {Martens}}, \
  and\ \bibinfo {author} {\bibfnamefont {J.-L.}\ \bibnamefont {Barrat}},\
  }\href {\doibase 10.1103/RevModPhys.90.045006} {\bibfield  {journal}
  {\bibinfo  {journal} {Rev. Mod. Phys.}\ }\textbf {\bibinfo {volume} {90}},\
  \bibinfo {pages} {045006} (\bibinfo {year} {2018})}\BibitemShut {NoStop}%
\bibitem [{\citenamefont {Durian}(1995)}]{Durian-1995}%
  \BibitemOpen
  \bibfield  {author} {\bibinfo {author} {\bibfnamefont {D.~J.}\ \bibnamefont
  {Durian}},\ }\href {\doibase 10.1103/PhysRevLett.75.4780} {\bibfield
  {journal} {\bibinfo  {journal} {Phys. Rev. Lett.}\ }\textbf {\bibinfo
  {volume} {75}},\ \bibinfo {pages} {4780} (\bibinfo {year}
  {1995})}\BibitemShut {NoStop}%
\bibitem [{\citenamefont {Basan}\ \emph {et~al.}(2009)\citenamefont {Basan},
  \citenamefont {Risler}, \citenamefont {Joanny}, \citenamefont
  {Sastre‐Garau},\ and\ \citenamefont {Prost}}]{Prost-2009}%
  \BibitemOpen
  \bibfield  {author} {\bibinfo {author} {\bibfnamefont {M.}~\bibnamefont
  {Basan}}, \bibinfo {author} {\bibfnamefont {T.}~\bibnamefont {Risler}},
  \bibinfo {author} {\bibfnamefont {J.}~\bibnamefont {Joanny}}, \bibinfo
  {author} {\bibfnamefont {X.}~\bibnamefont {Sastre‐Garau}}, \ and\ \bibinfo
  {author} {\bibfnamefont {J.}~\bibnamefont {Prost}},\ }\href {\doibase
  10.2976/1.3086732} {\bibfield  {journal} {\bibinfo  {journal} {HFSP Journal}\
  }\textbf {\bibinfo {volume} {3}},\ \bibinfo {pages} {265} (\bibinfo {year}
  {2009})},\ \bibinfo {note} {pMID: 20119483},\ \Eprint
  {http://arxiv.org/abs/https://doi.org/10.2976/1.3086732}
  {https://doi.org/10.2976/1.3086732} \BibitemShut {NoStop}%
\bibitem [{\citenamefont {Montel}\ \emph {et~al.}(2011)\citenamefont {Montel},
  \citenamefont {Delarue}, \citenamefont {Elgeti}, \citenamefont {Malaquin},
  \citenamefont {Basan}, \citenamefont {Risler}, \citenamefont {Cabane},
  \citenamefont {Vignjevic}, \citenamefont {Prost}, \citenamefont {Cappello},\
  and\ \citenamefont {Joanny}}]{Montel-2011}%
  \BibitemOpen
  \bibfield  {author} {\bibinfo {author} {\bibfnamefont {F.}~\bibnamefont
  {Montel}}, \bibinfo {author} {\bibfnamefont {M.}~\bibnamefont {Delarue}},
  \bibinfo {author} {\bibfnamefont {J.}~\bibnamefont {Elgeti}}, \bibinfo
  {author} {\bibfnamefont {L.}~\bibnamefont {Malaquin}}, \bibinfo {author}
  {\bibfnamefont {M.}~\bibnamefont {Basan}}, \bibinfo {author} {\bibfnamefont
  {T.}~\bibnamefont {Risler}}, \bibinfo {author} {\bibfnamefont
  {B.}~\bibnamefont {Cabane}}, \bibinfo {author} {\bibfnamefont
  {D.}~\bibnamefont {Vignjevic}}, \bibinfo {author} {\bibfnamefont
  {J.}~\bibnamefont {Prost}}, \bibinfo {author} {\bibfnamefont
  {G.}~\bibnamefont {Cappello}}, \ and\ \bibinfo {author} {\bibfnamefont {J.-F.
  m.~c.}\ \bibnamefont {Joanny}},\ }\href {\doibase
  10.1103/PhysRevLett.107.188102} {\bibfield  {journal} {\bibinfo  {journal}
  {Phys. Rev. Lett.}\ }\textbf {\bibinfo {volume} {107}},\ \bibinfo {pages}
  {188102} (\bibinfo {year} {2011})}\BibitemShut {NoStop}%
\bibitem [{\citenamefont {Malmi-Kakkada}\ \emph {et~al.}(2018)\citenamefont
  {Malmi-Kakkada}, \citenamefont {Li}, \citenamefont {Samanta}, \citenamefont
  {Sinha},\ and\ \citenamefont {Thirumalai}}]{Thirumalai-2018}%
  \BibitemOpen
  \bibfield  {author} {\bibinfo {author} {\bibfnamefont {A.~N.}\ \bibnamefont
  {Malmi-Kakkada}}, \bibinfo {author} {\bibfnamefont {X.}~\bibnamefont {Li}},
  \bibinfo {author} {\bibfnamefont {H.~S.}\ \bibnamefont {Samanta}}, \bibinfo
  {author} {\bibfnamefont {S.}~\bibnamefont {Sinha}}, \ and\ \bibinfo {author}
  {\bibfnamefont {D.}~\bibnamefont {Thirumalai}},\ }\href {\doibase
  10.1103/PhysRevX.8.021025} {\bibfield  {journal} {\bibinfo  {journal} {Phys.
  Rev. X}\ }\textbf {\bibinfo {volume} {8}},\ \bibinfo {pages} {021025}
  (\bibinfo {year} {2018})}\BibitemShut {NoStop}%
\bibitem [{\citenamefont {Angelini}\ \emph {et~al.}(2011)\citenamefont
  {Angelini}, \citenamefont {Hannezo}, \citenamefont {Trepat}, \citenamefont
  {Marquez}, \citenamefont {Fredberg},\ and\ \citenamefont
  {Weitz}}]{Angelini-2011}%
  \BibitemOpen
  \bibfield  {author} {\bibinfo {author} {\bibfnamefont {T.~E.}\ \bibnamefont
  {Angelini}}, \bibinfo {author} {\bibfnamefont {E.}~\bibnamefont {Hannezo}},
  \bibinfo {author} {\bibfnamefont {X.}~\bibnamefont {Trepat}}, \bibinfo
  {author} {\bibfnamefont {M.}~\bibnamefont {Marquez}}, \bibinfo {author}
  {\bibfnamefont {J.~J.}\ \bibnamefont {Fredberg}}, \ and\ \bibinfo {author}
  {\bibfnamefont {D.~A.}\ \bibnamefont {Weitz}},\ }\href {\doibase
  10.1073/pnas.1010059108} {\bibfield  {journal} {\bibinfo  {journal}
  {Proceedings of the National Academy of Sciences}\ }\textbf {\bibinfo
  {volume} {108}},\ \bibinfo {pages} {4714} (\bibinfo {year} {2011})},\ \Eprint
  {http://arxiv.org/abs/https://www.pnas.org/content/108/12/4714.full.pdf}
  {https://www.pnas.org/content/108/12/4714.full.pdf} \BibitemShut {NoStop}%
\bibitem [{\citenamefont {Poincloux}\ \emph {et~al.}(2011)\citenamefont
  {Poincloux}, \citenamefont {Collin}, \citenamefont {Liz{\'a}rraga},
  \citenamefont {Romao}, \citenamefont {Debray}, \citenamefont {Piel},\ and\
  \citenamefont {Chavrier}}]{Poincloux-2011}%
  \BibitemOpen
  \bibfield  {author} {\bibinfo {author} {\bibfnamefont {R.}~\bibnamefont
  {Poincloux}}, \bibinfo {author} {\bibfnamefont {O.}~\bibnamefont {Collin}},
  \bibinfo {author} {\bibfnamefont {F.}~\bibnamefont {Liz{\'a}rraga}}, \bibinfo
  {author} {\bibfnamefont {M.}~\bibnamefont {Romao}}, \bibinfo {author}
  {\bibfnamefont {M.}~\bibnamefont {Debray}}, \bibinfo {author} {\bibfnamefont
  {M.}~\bibnamefont {Piel}}, \ and\ \bibinfo {author} {\bibfnamefont
  {P.}~\bibnamefont {Chavrier}},\ }\href {\doibase 10.1073/pnas.1010396108}
  {\bibfield  {journal} {\bibinfo  {journal} {Proceedings of the National
  Academy of Sciences}\ }\textbf {\bibinfo {volume} {108}},\ \bibinfo {pages}
  {1943} (\bibinfo {year} {2011})},\ \Eprint
  {http://arxiv.org/abs/https://www.pnas.org/content/108/5/1943.full.pdf}
  {https://www.pnas.org/content/108/5/1943.full.pdf} \BibitemShut {NoStop}%
\bibitem [{\citenamefont {Dell'Arciprete}\ \emph {et~al.}(2018)\citenamefont
  {Dell'Arciprete}, \citenamefont {Blow}, \citenamefont {Brown}, \citenamefont
  {Farrell}, \citenamefont {Lintuvuori}, \citenamefont {McVey}, \citenamefont
  {Marenduzzo},\ and\ \citenamefont {Poon}}]{DellArciprete-2018}%
  \BibitemOpen
  \bibfield  {author} {\bibinfo {author} {\bibfnamefont {D.}~\bibnamefont
  {Dell'Arciprete}}, \bibinfo {author} {\bibfnamefont {M.~L.}\ \bibnamefont
  {Blow}}, \bibinfo {author} {\bibfnamefont {A.~T.}\ \bibnamefont {Brown}},
  \bibinfo {author} {\bibfnamefont {F.~D.~C.}\ \bibnamefont {Farrell}},
  \bibinfo {author} {\bibfnamefont {J.~S.}\ \bibnamefont {Lintuvuori}},
  \bibinfo {author} {\bibfnamefont {A.~F.}\ \bibnamefont {McVey}}, \bibinfo
  {author} {\bibfnamefont {D.}~\bibnamefont {Marenduzzo}}, \ and\ \bibinfo
  {author} {\bibfnamefont {W.~C.~K.}\ \bibnamefont {Poon}},\ }\href {\doibase
  10.1038/s41467-018-06370-3} {\bibfield  {journal} {\bibinfo  {journal} {Nat.
  Comms.}\ }\textbf {\bibinfo {volume} {9}},\ \bibinfo {pages} {1} (\bibinfo
  {year} {2018})}\BibitemShut {NoStop}%
\end{thebibliography}%
\bibliographystyle{apsrev4-1} %the RSC's .bst file

\end{document}